\documentclass[10pt,aps,floats,prl,
twocolumn,showpacs]{revtex4}
\usepackage{graphicx}
\usepackage{epsfig}
\newcommand{\nn}{\nonumber}
\newcommand{\beq}{\begin{equation}}
\newcommand{\eeq}{\end{equation}}
\newcommand{\be}{\begin{eqnarray}}
\newcommand{\ee}{\end{eqnarray}}

\def\+{\dagger}

\begin{document}
\title{511~KeV Photons From Color Superconducting Dark Matter}
\author{David H. Oaknin and Ariel R.Zhitnitsky}

\affiliation{ Department of Physics and Astronomy, University of British 
Columbia, Vancouver, BC, V6T 1Z1, CANADA}

\begin{abstract}
We discuss the possibility that the recent detection of $511 $  keV $ \gamma $
rays from the galactic bulge, as observed by INTEGRAL, can be naturally explained by 
 the   supermassive very dense droplets (strangelets)  of dark matter. These droplets
 are assumed to be made of ordinary light 
quarks (or antiquarks)  condensed in non-hadronic color superconducting 
phase.  The droplets   can carry electrons (or positrons) 
 in the bulk or/and on the  surface. 
 The  $e^+e^-$ annihilation events   take place
due to the collisions of electrons from the visible matter with positrons
from dark matter droplets which  may result  in the  bright 
$511$~KeV $\gamma$-ray line from the bulge of the Galaxy.  

\end{abstract}
\maketitle

{\em Introduction}.---%
The recent detection by the SPI spectrometer on the International 
Gamma-Ray Astrophysics Laboratory~(INTEGRAL) satellite of a bright 
$511$~KeV $\gamma$-ray line from the bulge of the Galaxy 
with spherically symmetric distribution \cite{Knodlseder:2003sv} has 
stirred the research of the fundamental physics that describes the 
cosmological dark matter.

The flux of $511$~KeV photons (with a width of about $3$~KeV), produced by 
thermalized electron positron pair annihilation processes, has been 
measured to be $9.9^{+4.7}_{-2.1} \times  
10^{-4}$~photons~$\mbox{cm}^{-2}$~$\mbox{s}^{-1}$ 
and has an angular distribution with half maximum 
at $9^o$ ($6^o$ to $18^o$ at $2\sigma$ confidence), in good 
agreement with previous measurements \cite{Milne:2001zs}.

The source of these thermalized positrons in the bulge of the Galaxy have 
been the subject of much debate. Some proposals suggest astrophysical 
processes, including neutron stars or black holes \cite{Ramaty}, pulsars 
\cite{Sturrock:zc}, radiactive nuclei from supernova \cite{Ramaty:pv} or 
cosmic ray interactions with the interstellar medium \cite{Ramaty:pv2}, 
but it is rather uncertain which fraction of positrons produced in such 
processes can escape and, moreover, how they could fill the whole galactic 
bulge \cite{Hooper:2003sh}.

Recently it has been discussed that light dark matter particles 
annihilating into $e^{+} e^{-}$ pairs in the galactic bulge may be the 
source of the thermalized positrons that produce the $511$~KeV 
emission line \cite{Boehm:2003bt}, see also
related works, \cite{Picciotto:2004rp}, \cite{Hooper:2004qf},  \cite{Casse:2004gw}.
The shallow density distribution of 
dark matter in the bulge of the galaxy $\rho(r) \sim r^{-\gamma}$, with 
$\gamma = 0.4$ to $0.8$, explains very naturally the angular distribution 
of detected $511$~KeV $\gamma$ photons. 
%

{\em Dark matter as color superconductor}.---%
We want to elaborate the proposal \cite{Boehm:2003bt} in the context of a 
cosmological scenario when dark matter consists of very dense 
(few times the nuclear density) macroscopic 
droplets of ordinary light quarks (  or/and  antiquarks\cite{qcdball},
\cite{BDMQCD} ) condensed in  
non-hadronic color superconducting phase, similar 
to the Witten's strangelets \cite{Witten:1984rs}.

In this Letter we will argue that color superconducting dark matter also 
provides a natural and simple framework to explain the detected emission 
of $511$~KeV photons from the galactic bulge with the appropriate angular 
distribution and intensity. Indeed, the main required ingredients of the proposal
are automatically present in our scenario: a large number of positrons is always
present in antimatter dark matter droplets, see below. 
We argued in \cite{BDMQCD} 
   that chunks of quarks or 
antiquarks in condensed color superconducting phase may be formed 
during the QCD phase transition and they may serve as dark matter.  
This scenario is based on the idea that while the universe is globally symmetric,
the antibaryon charge can be stored in 
chunks of dense color superconducting (CS) antimatter. In different words, 
the baryon asymmetry of the universe may not necessarily be expressed as 
a net baryon number if the anti-baryon charge 
is accumulated in form of the diquark condensate in CS phase, rather than in form of 
free anti-baryons in hadronic phase. 
We 
explained  in  \cite{BDMQCD}  why such a scenario does not contradict the current observational 
data on antimatter in the Universe. This is  mainly due to the very small volume
occupied by dense droplets and specific features of interaction between
color superconducting phase and conventional hadronic  matter.
 We also argued that 
 the observed cosmological ratio between the energy densities
of dark and baryonic matter, $\Omega_{DM}\sim\Omega_{B}$ within an 
order of  magnitude, finds its natural explanation in this scenario: both  
contributions to $\Omega$ originated from the same physics at the same 
instant during the QCD phase transition. As is known, 
the relation $\Omega_B\sim \Omega_{DM}$ between the two very different contributions
to $\Omega$ is extremely difficult to explain in models that invoke a   DM candidates
not related to the ordinary quark/baryon degrees of freedom.
The baryon to entropy ratio $n_{B}/n_{\gamma}\sim 10^{-10}$ would also
be a natural outcome in this scenario. 
We refer to the original papers   \cite{qcdball}, \cite{BDMQCD}
for the details. Here we want to mention only  the fact  that
the  baryon charge of massive droplets  does not change the 
nucleosynthesis calculations because in the color superconducting   phase it 
is not available for nuclearsynthesis 
when the baryon charge is locked in the coherent superposition
of Cooper pairs. Therefore, while 
the massive droplets carry a large baryon charge, they do not contribute
to  $\Omega_B$, but rather, they    do contribute to the 
``non-baryonic" cold dark matter $\Omega_{DM}$ of the universe \cite{qcdball}, \cite{BDMQCD}.
 
Before we estimate the probability of the $e^+e^-$ annihilation which results in $511$~KeV line,
we would like to make a short review on basic properties of dense droplets 
in color superconducting phase, which will be referred as QCD balls  \cite{qcdball},\cite{BDMQCD}
 in what follows.
The color superconducting  state  of quark matter is a novel phase in QCD that 
is realized when light quarks are squeezed to a density which is 
a few times the nuclear density.  The ground state  in this phase is a single coherent state with  
diquark condensation, analogous to Cooper pairs of electrons in 
BCS theory of ordinary superconductors. In the approximation of three 
light  quarks  $m_u, m_d, m_s \ll \mu$ and relatively large chemical potential $\mu\gg \Lambda_{QCD}$,
 the so-called  color-flavour-locking~(CFL) phase is a
preferred state of matter, see original papers \cite {cs_n} and 
recent review \cite{cs_r} on the subject.  For physical value of $m_s$ and 
$\mu\simeq 500 $ MeV a number of different CS phases may  result. It is not the goal of  this letter
to describe a variety of possibilities when  parameters (such as $m_s$ and $\mu$) vary.
 Rather, we want to emphasize below that the sufficient number of positrons 
  will always accompany the QCD balls made of antimatter (QCD anti-balls).

Indeed, first of all, consider the most symmetric, the CFL phase. While this phase
does not support the leptons in the bulk \cite{Rajagopal:2000ff}, the finite volume effects 
lead to the accumulation of the positive charge on the surface  \cite{Madsen:2001fu},
which must be neutralized by negative electron charge (for droplets made of matter). 
For droplets made of antimatter, the corresponding positron charge will be accumulated. 
In most other phases which may be realized in nature,
 the leptons will be present on the surface as well as  in the bulk of a droplet.
 The electron density   can be roughly  estimated as $n_e\simeq \frac{\mu_e^3}{3\pi^2}$,
 with $\mu_e$ being the electron chemical potential (in case of matter droplets) or positron chemical potential in case of anti-matter droplets. In this case, the electrons (positrons) in droplets
 can be treated as fermi liquid.
A numerical estimation of 
$\mu_e$  strongly depends on the specific details  of CS phase under consideration,
 and varies  from few $MeV$ to tens $MeV$,~\cite{Alford:2002kj}- \cite{olinto}. 
 
 However,
  the   important property which plays essential role 
 for the present work  (and which is shared by all CS phases),  
  is as follows.  
  Consider an  electron which hits the DM droplet (made of anti matter).
  What is the fate of this non relativistic   electron?
  It can form a bound state (positronium with arbitrary quantum numbers $|n,l,m\rangle$)
    which eventually decays to  two $\sim 511 KeV $ photons.  
  It may also annihilate with energetic positron into two photons in non  resonance manner with emitting 
  2 $\gamma$'s with a typical energy determined by $\mu_e$ ( few $MeV$ scale).  However, 
  the probability for the later annihilation is suppressed by small coupling constant $\alpha^2$,
  in comparison with the former process, when 
   the probability for the  formation of positronium from  two nonrelativistic
   particles $e^+$ and $e^-$ could be order of one.
  Indeed, the probability for the positronium formation (as well as for its decay to free $e^+e^-$ pair)
  if the system gets an instantaneous jolt (with relative momentum $q=mv$) is determined
  by the overlap of two wave functions $ \sim |\langle\psi_{out}|\psi_{in}\rangle|^2 \sim
  |\int e^{-r/a}e^{i\vec{q}\vec{r}}d^3r|^2$ where $e^{-r/a}$ represents a typical positronium wave function
  in state  $|n,l,m\rangle$ with $a\simeq 10^{-8}cm$. Of course, this expression assumes the validity of instantaneous
  perturbation theory when parameter $qa >> 1$, while the maximum probability is achieved when
  $qa \simeq 1$, see below.  It is obvious that the main contribution
  to the positronium formation 
  is due to the process when the incoming electron  picks  up a positron from the droplet
  with a typical velocity determined by the condition: $qa\sim 1$. This corresponds to $v/c\sim \alpha$
 for a typical positronium size, $a\sim{\hbar^2}/{me^2}$.
  Eventually, it 
  decays to two $\sim 511 KeV$  photons.  The flux of emitted photons 
produced by this mechanism will naturally have a width of order $\Gamma/(511 KeV)\sim v/c \sim \alpha\sim 10^{-2}$, which is what observations apparently suggest  \cite{Knodlseder:2003sv}. 
  We note that the positronium formation (with consequent emission  of $511 KeV$  photons)
  is expected to occur on the surface of the droplet  such that considerable portion of 
  $511 KeV$  photons leave the system without reabsorption.  
   
  To conclude:
  the annihilation cross section for  the electron falling to the DM 
  anti-droplet is given by the geometrical size
  of the object, $4\pi R^2$, while a typical width
  of  outgoing flux of $511 KeV$ photons is of order $\Gamma\sim \alpha m\sim $ few $KeV$.
  These features are very universal, do not depend on specific details of the phase under consideration,  and remain unaltered  for all possible CS phases.
  With these remarks in mind, we estimate the $e^+e^-$ annihilation rate 
and the flux of $511 $ KeV photons and compare it with the 
observational available data. 
  
{\em First rough estimate  }---%
We start with a first estimation of the annihilation assuming that
visible matter density follows the spatial distribution of dark 
matter, with the fixed ratio given by the cosmological 
$\Omega_B/\Omega_{DM}$. We also assume that the  electron density from 
the visible matter is roughly determined by  the number density of 
protons.  The system could be in ionized state (HII) or in neutral atomic hydrogen state.
 It is 
quite obvious that corresponding calculations would  lead to   strong
 underestimation of the annihilation rate because 
 the visible matter is strongly peaked in the center of galaxy, the effect
 which is completely ignored in our first estimate. The positive elements
of  such an assumptions are: a) it allows us to follow closely the original 
analysis in \cite{Boehm:2003bt}, such that the spatial 
integration over matter density can be extracted from this paper, and 
the  corresponding comparison with ~ \cite{Boehm:2003bt} can be made;  
b) it gives us a lower bound for the corresponding annihilation rate as 
argued above. More importantly, this lower bound depends only on a 
typical size of the droplets, and does not depend on specific assumptions on 
behavior of visible matter density in the center of galaxy. 
 

The estimation of the flux of $511$~KeV photons coming to Earth from the bulge of the 
Galaxy along the angular direction $\Omega$  goes as follows.
As we mentioned above,  the number of electrons 
is roughly determined by  the number density of 
protons, $n_{e^-}\simeq n_B$, and all electrons which hit the QCD anti-ball (antidroplet made of antimatter) with radius $R$ will annihilate such that a considerable portion 
of the process will lead to the production 
  of two  $511$~KeV photons. The probability per unit time $\frac{dW}{dt}$ that this happens in the presence of a single QCD ball is given by
\beq
\label{1}
\frac{dW}{dt}=4\pi R^2n_{e^-}v  \simeq 4\pi R^2n_{B}v \simeq 
 4\pi R^2\frac{0.15\rho_{DM}}{1 GeV}v ,
\eeq
where $v/c\sim 10^{-3}$ and we express the baryon density
in terms of dark matter density, $ 1 GeV\cdot n_B\simeq\rho_B\simeq \Omega_B/\Omega_{DM}
\rho_{DM} \simeq 0.15\rho_{DM}$ to make  our first  rough estimate.
In order to estimate the probability of such events per unit volume per unit time $\frac{dW}{dVdt}$
one should multiply  eq.(\ref{1}) by the inverse volume occupied by a typical QCD ball with a typical baryon charge $B$. In our framework when the dark matter is identified with
QCD balls and anti balls with typical mass $M\simeq m_N\cdot B$, the corresponding
number density of the DM particles is nothing but $n_{DM} \simeq \frac{\rho_{DM}}{1 GeV }\frac{1}{B}$.
Therefore, we arrive to the following estimate,
\beq
\label{2}
\frac{dW}{dVdt}\simeq 
0.15\cdot v \cdot \frac{4\pi R^2}{B} \cdot (\frac{\rho_{DM}}{1 GeV})^2 .
\eeq
The total flux of photons resulting from annihilation is obtained by 
integrating eq. (\ref{2}) over the line of sight and over the whole 
solid angle of observation. The numerical evaluation was done in 
\cite{Boehm:2003bt}. We follow their analysis and implement it in our
framework. We arrive at
\be
\label{flux}
\Phi = \int ds \int_{\Delta \Omega} d\Omega\frac{dW}{dVdt} \simeq  \nn \\
10^{-3}cm^{-2}s^{-1}\cdot {\bar J}(\Delta \Omega) \Delta \Omega\cdot 
  \left(\frac{10^{18}}{B}\right)^{1/3} 
  \ee
where ${\bar J}(\Delta \Omega) \Delta \Omega \equiv \int_{\Delta \Omega} 
d\Omega J(\Omega)$ with
\begin{equation}
J(\Omega) = (\frac{1}{0.3 GeV/cm^3})^2\frac{1}{8.5kpc}\int ds\left[\rho_{DM}(s)\right]^2.
\end{equation}
In expression  (\ref{flux}) we traded $R$ from eq. (\ref{2})  in favor of $B\simeq 
\frac{4\pi R^3}{3} n_{CS}$  assuming that a typical baryon number density
in  color superconducting phase,   $n_{CS}$,  is  three times the nuclear saturation density, 
$n_{CS} \simeq 3 n_0$ with $n_0\sim (108 MeV)^3$.

The factor $J(\Omega)$ has been evaluated in reference \cite{Boehm:2003bt} 
for different density profiles $\xi(r) \propto r^{-\gamma}$ 
with   $\gamma = 0.4 - 0.8$ providing the best fit.
For these favorite $\gamma's$ the value ${\bar J}(\Delta \Omega)\Delta \Omega$
 has been shown to vary    
in the range $ 0.3 - 1.6$. This value should be substituted into eq. (\ref{flux})    
and compared with the observations,  $9.9^{+4.7}_{-2.1} \times  
10^{-4}$~photons~$\mbox{cm}^{-2}$~$\mbox{s}^{-1}$. 

As we mentioned above, we consider this estimate as the lowest extreme case 
(within our framework).  Indeed,   our assumption on visible matter density distribution,
$\rho_B\simeq  
 0.15\rho_{DM}$ with $\rho_{DM}\sim r^{-\gamma}$ and $\gamma=0.6$,
 normalized to the local density  $ \rho_{DM}\simeq 0.3 GeV/cm^3$ would lead 
 to the total visible material (within $8.5$ kpc region ) of  about $4 \cdot 10^9 M_{\odot} $ 
 instead of observed $ \sim 10^{11} M_{\odot} $.

Nevertheless this simple estimate  is very instructive. First of all, 
one can explicitly compare our expression (\ref{flux}) with the corresponding
formula from ref. \cite{Boehm:2003bt} when  the same factor describing DM distribution,
${\bar J}(\Delta \Omega) 
\Delta \Omega   $ enters  the relevant formulae.
    Secondly, even the obviously underestimated expression (\ref{flux}) is  not 
in contradiction with    the existing
bound on such kind of dense droplets, see eq. (20) in ref.\cite{qcdball}
where  bound  $B> 10^{20}$ is quoted.
  
  {\em  Unaccounted  effects. Further complications }---%
  Here we want to discuss some new  effects (ignored above)
  which certainly increase the rate.
  Unfortunately,   the corresponding estimates  are strongly 
  model dependent, see below, and, therefore, should be taken  with  some cautious. 
First,   let us   take into account  the properties of the 
visible matter distribution in the galaxy in a more appropriate way  than it is done above.
  We replace formula (\ref{2}) by the following expression,
\be
\label{5}
\label{p}
\frac{dW}{dVdt}(r) \simeq 
  \frac{4\pi R^2}{B}   \cdot v  \cdot (\frac{\rho_{B}}{1 GeV}) \cdot  (\frac{\rho_{DM}}{1 GeV}),
\ee
The number density of electrons and  the number density  of dark 
matter particles are estimated as before, 
  $n_{e^-} \sim n_B  \simeq(\frac{\rho_{B}}{1 GeV}) ,~ 
 n_{DM} \simeq \frac{\rho_{DM}}{1 GeV }\frac{1}{B}$.
We parameterize  DM density as  
 \be
\label{DM}
 \rho_{DM}\simeq 0.03\frac{M_{\odot}}{pc^3}\frac{1}{(r/kpc)^{0.6}},
 \ee
normalized to the local density $\rho_{DM}\simeq 0.3 GeV/cm^3$, 
which is the central value adopted by \cite{Boehm:2003bt}.
For the visible matter we adopt the following
scaling behavior  ( close to the $r^{-2}$ behavior of an izothermal sphere\cite{book}),
  \be
\label{B}
  \rho_{B}\simeq 0.7\frac{M_{\odot}}{pc^3}\frac{1}{(r/kpc)^{1.8}},
 \ee 
normalized to the total visible mass  of $M_{tot}=\int^{8.5  kpc}d^3x 
\rho_{B}\simeq10^{11}M_{\odot}$ within $8.5 kpc$.
 We notice that such a peaked distribution of visible matter would, in 
principle, produce a narrower distribution of 511 KeV photons 
  than currently preferred by observational values, 
$\frac{dW}{dVdt}(r) \sim r^{-2\gamma}$, with $\gamma$ between $0.4$ and 
$0.8$ ~ \cite{Boehm:2003bt}. However, if  we take $\gamma\simeq 0$ for the dark matter,
the angular distribution which follows from eq. (\ref{5},\ref{B}) would be close 
to the upper bound of the preferred value\cite{Boehm:2003bt}.
 
 Combining eqs. (\ref{p},\ref{DM},\ref{B}) we arrive to the following 
final expression for the flux
\be
\label{flux1}
\Phi = \int  dr  \Delta \Omega \frac{dW}{dVdt} \simeq   
10^{-3}   \mbox{cm}^{-2}\mbox{s}^{-1}  \left(\frac{10^{33}}{B}\right)^{1/3},
  \ee
 In obtaining the estimate (\ref{flux1})
we cut off the integral 
$\int^{8.5kpc}_{0.5pc} dr$ at  $0.5 pc$ at small distances where the visible matter rises very fast
$\sim r^{-2.7}$ while DM behavior at such scales is absolutely unknown. Such a cut off obviously brings
a large uncertainty into our estimate. There is also large uncertainty
due to the unknown scaling properties of the dark matter distribution at small distances.
Finally, different clumps and structures (such as stars, MACHOS, astreroids, etc)   of the baryonic matter can strongly enhance
 the estimate (\ref{flux1}) due to the fact
 that a large number of positrons from the bulk (rather than from the surface)
  of the QCD balls can participate in annihilation. 
   Unfortunately, we do not know how to account this effect properly.
 The main goal here is  to demonstrate    the 
 sensitivity of the calculations   with variation of   the visible and DM matter  distributions: the
difference between two estimates, (\ref{flux}) and (\ref{flux1}) is almost  5 orders of magnitude.


{\em Conclusion  }---%
    The main goal of the present letter is to argue 
  that the color superconducting dark matter ( introduced with quite  different
motivation~\cite{qcdball}, \cite{BDMQCD})
  provides a natural and simple framework to explain the detected emission 
of $511$~KeV photons from the galactic bulge with the appropriate angular 
distribution and intensity.  While there are many other possibilities to explain this
rate   based on some specific DM features, such as annihilation or decay,
the present proposal   is unique in many respects and can be easily
discriminated from other explanations based on DM particles. Indeed, a unique
feature of our scenario is proportionality 
of the local flux of photons to both the density of 
visible as well as dark matter, see eq.(\ref{p}). In other DM based 
explanations the local flux does depend only on the distribution of dark 
matter. 
 The corresponding matter distributions are obviously  very different. In particular,
 an observation of the effect on the same level
 but in a different  direction  (not pointing to the center of the galaxy)
 would rule out our explanation.  
 
 We   also point out that  
 $\bar{q}q$ annihilation
 might be sufficiently large for relatively energetic protons (with kinetic energy
 about $1 GeV$)   \cite{BDMQCD}.
  In this case $e^+e^-$ annihilation with single bright $511 KeV$ line
 (discussed in the present paper) would be accompanied by the wide (70 MeV -1 GeV) 
  $\gamma$  spectral density due to the 
   baryon- antibaryon  annihilation. These very different spectra in different frequency regions must be related to each other    due to their  common origin.  
  Corresponding calculations are beyond the scope of the present work;
  however, a very simplified estimate of the corresponding flux 
 can be obtained by replacing electron velocity $v$
  in formula (\ref{5}) by a proton velocity $v_p/v\sim \sqrt{m_e/m_p}\sim 2 \cdot 10^{-2}$
  \cite{footnote}.
   This  corresponds to the assumption
 of the thermal equilibrium between electrons and protons 
 in the ionized region  in the bulge of the galaxy  (the HII has a vertical scale hight of  $\sim 90$ pc
 \cite{book}).      
Estimated in such a way flux
 is definitely not in immediate contradiction with observations, where some access of $\gamma$ rays  indeed has been observed by EGRET.
   We should add that the observed access has been interpreted
  in \cite{cesarini}, \cite{deboer} as due to the dark matter annihilation, and  in \cite{khlopov} as 
  due to $p\bar{p}$ annihilation. One more phenomenological consequence of the suggested 
  scenario  is that  baryon- antibaryon  annihilation which always accompany 
  $511 KeV$ line  eventually may be responsible for a "non observation"
of the cusp behavior near the Galactic Center. It might be 
worthwhile to investigate this possibility in more details  in future.   

 {\em Acknowledgments :}  
ARZ is  thankful to   Alex Vilenkin and other participants  of the joint Tufts/CfA/MIT Cosmology seminar,
and DHO is thankful to the participants of the seminar at Weizmann Institute
(where this and accompanying works\cite{qcdball},\cite{BDMQCD}  were presented)
 for the discussions and critical remarks. ARZ also thanks Lev Kofman, Slava Mukhanov, Jes Madsen,  Frank Wilczek  and Jeremy Heyl for  discussions.
This work was supported in part by the National Science and Engineering
Research Council of Canada. 
  

\begin{references}
\bibitem{footnote} In such a rough
   estimate we assume
  that  transmission coefficient for a proton hitting the ball is order of one. As we argued in \cite{BDMQCD} the transmission can be actually  strongly suppressed for the small energies 
  due to the large reflection from the superconducting
  region of the  bulk of the QCD ball.
\end{references}
\end{document}